\documentclass[traditabstract]{aa}  
\usepackage{txfonts}
\usepackage{epstopdf}

\usepackage{graphicx,longtable,lscape,natbib,amssymb,amssymb,amsmath}
\bibpunct{(}{)}{;}{a}{}{,} 
\newcommand{\ltsima} {$\; \buildrel < \over \sim \;$}  
\newcommand{\gtsima} {$\; \buildrel > \over \sim \;$}  
\newcommand{\lta} {\lower.5ex\hbox{\ltsima}}  
\newcommand{\gta} {\lower.5ex\hbox{\gtsima}}  
\newcommand{\ha} {H$\alpha$}  
\newcommand{\hb} {H$\beta$}  

\newcommand{\ergs}{\>{\rm erg}\,{\rm s}^{-1}}
\newcommand{\ergshz}{\>{\rm erg}\,{\rm s}^{-1}\,{\rm Hz}^{-1}}

\newcommand{\kms}{$\rm{\,km \,s}^{-1}$}

\newcommand{\forb}[2]{\mbox{$[{\rm #1\, #2}]$}}
\newcommand{\oiii}{\forb{O}{III}}
\newcommand{\oi}{\forb{O}{I}\,}
\newcommand{\sii}{\forb{S}{II}\,}
\newcommand{\nii}{\forb{N}{II}\,}

\usepackage{color}

\begin{document}

\title{The MURALES survey. I.} \subtitle{A dual AGN in the radio galaxy 3C~459?}

\author{Barbara Balmaverde\inst{1} 
                \and Alessandro Capetti\inst{2}
                \and Alessandro Marconi\inst{3,4}
                \and Giacomo Venturi\inst{3,4}
                \and M. Chiaberge\inst{5,6}
                \and R.D. Baldi\inst{7}
                \and S. Baum\inst{10,15}
                \and R. Gilli\inst{8}
                \and P. Grandi\inst{9}
                \and E. Meyer\inst{13}
                \and G. Miley\inst{11}
                \and C. O$'$Dea\inst{10,14}
                \and W. Sparks\inst{5}
                \and E. Torresi\inst{9} 
                \and G. Tremblay\inst{12}}
\institute {INAF - Osservatorio Astronomico di Brera, via E. Bianchi 46, 23807, Merate, Italy
  \and INAF - Osservatorio Astrofisico di Torino, Via Osservatorio 20, I-10025 Pino Torinese, Italy
\and Dipartimento di Fisica e Astronomia, Universit\`a di Firenze, via G. Sansone 1, 50019 Sesto Fiorentino (Firenze), Italy
 \and INAF - Osservatorio Astrofisico di Arcetri, Largo Enrico Fermi 5, I-50125 Firenze,Italy
 \and Space Telescope Science Institute, 3700 San Martin Dr., Baltimore, MD 21210, USA
\and Johns Hopkins University, 3400 N. Charles Street, Baltimore, MD 21218, USA
 \and Department of Physics and Astronomy, University of Southampton, Highfield, SO17 1BJ, UK
 \and INAF - Osservatorio Astronomico di Bologna, Via Gobetti 93/3, I-40129 Bologna, Italy
\and INAF - IASFBO, Via Gobetti 101, I-40129, Bologna, Italy
\and Department of Physics and Astronomy, University of Manitoba, Winnipeg, MB R3T 2N2, Canada
 \and Leiden Observatory, Leiden University, PO Box 9513, NL-2300 RA, Leiden, the Netherlands
\and Department of Physics and Yale Center for Astronomy \& Astrophysics, Yale University, 217 Prospect Street, New Haven, CT 06511, USA
\and University of Maryland Baltimore County, 1000 Hilltop Circle, Baltimore, MD 21250, USA
\and School of Physics \& Astronomy, Rochester Institute of Technology, Rochester, NY 14623
\and Carlson Center for Imaging Science, Rochester Institute of Technology, Rochester, NY 14623
}
\offprints{balmaverde@oato.inaf.it} 

\date{} 

\abstract{We observed the FR~II radio galaxy 3C~459 (z=0.22) with the MUSE
  spectrograph at the Very Large Telescope (VLT) as part of the MURALES project (a MUse RAdio Loud
  Emission line Snapshot survey). We detected diffuse nuclear emission and a
  filamentary ionized gas structure forming a one-sided, triangular-shaped
  region extending out to $\sim$80 kpc. The central emission line region is
  dominated by two compact knots of similar flux: the first (N1) cospatial
  with the radio core and the (N2) second located 1$\farcs$2 (5.3 kpc) to the
  SE. The two regions differ dramatically from the point of view of velocity
  (with an offset of $\sim$400 km/s), line widths, and line ratios. This
  suggests that we are observing a dual AGN system formed by a radio loud AGN
  and type 2 QSO companion, which is  the result of the recent merger that also produced
  its disturbed host morphology. The alternative possibility that N2 is just a
  bright emission line knot resulting from, for example, a jet-cloud interaction, is
  disfavored because of 1) the presence of a high ionization bicone whose apex
  is located at N2; 2) the observed narrow line widths; 3) its line luminosity
  ($\sim10^{42}$erg s$^{-1}$) typical of luminous QSOs; and 4) its location,
  which is offset from the jet path. The putative secondary AGN must be highly
  obscured, since we do not detect any emission in the Chandra and infrared
  Hubble Space Telescope images.  } 

\titlerunning{A dual AGN in 3C~459?} 
\authorrunning{B. Balmaverde et al.}
 \maketitle

\section{Introduction}
\label{intro}

Mergers play a fundamental role in the hierarchical models of galaxy formation
and evolutions. Among other effects, mergers with a gas-rich galaxy funnel gas
 toward the central galactic regions \citep{mihos96}, potentially
enhancing star formation and triggering the Active Galactic Nuclei (AGN) activity (e.g.,
\citealt{treister12,hopkins14}).

Since most, if not all, galaxies contain a super massive black hole (SMBH)  in their center, a merging of
two galaxies might lead to the formation of a black hole binary. If both
galaxies are accreting they could shine as a AGN pair that evolves on a $\sim$100
Myr timescale \citep{begelman80} into a dual AGN before the final
coalescence. In this intermediate evolutionary stage dual AGNs are separated
by less than 10 kpc and, according to models, the AGN activity and star
formation are the most vigorous (e.g., \citealt{vanwassenhove12,
  blecha13}). They can therefore be  identified with X-ray (e.g.,
\citealt{koss12}) or optical spectroscopic observations (e.g.,
\citealt{muller15,smith10,wang09}).  Hundreds of AGN pairs with $>$10 kpc
separations have been discovered so far \citep{myers08,hennawi10}. However,
due to various observational difficulties, there are only a few confirmed
kiloparsec-scale dual AGNs. Therefore, despite the effort in recent decades,
binary AGN turned out to be rather elusive objects (e.g.,
\citealt{koss12,rosario11,fu11}). This is likely, in part, because
AGN in galaxy mergers are likely more obscured than those in isolated galaxies
\citep{kocevski15,koss16,ricci17,satyapal17}. In the mid-infrared dust
extinction is attenuated and the selection of candidates using mid-IR colors
from the Wide Field Infrared Survey Explorer (WISE) improves the success rate
for dual AGN confirmation \citep{satyapal17,ellison17}. 
In the radio band, 
it is possible to resolve binary AGN at an angular distance 0.1-1 milli-arcseconds.
All the binary black holes separated by less than 10 parsec have been confirmed 
with this method, (e.g.,  \citealt{rodriguez06,gabanyi16,kharb17}). Long-term monitoring programs of AGN with the VLBI,
such as the MOJAVE project (Monitoring Of Jets in Active galactic nuclei with VLBA Experiments),
are investigating the kinematics of the radio jets on parsec scale.
Many AGN jets undergo significant changes in the jet projected direction \citep{lister13}
and in some cases they show evidence of long-term precession  \citep{alexander85,steenbrugge08}
possibly caused by a companion.
However, a nondetection in the radio band does not exclude the presence of dual AGN, since
only $\sim$10\% of the nuclei are radio loud.

Radio galaxies are among the best sources for a search of a dual/binary AGN.
In fact, a major merger between two galaxies of similar mass ($\sim10^{10}
-10^{11}$ M$_\odot$; e.g., Best et al. 2005) hosting high-mass SMBHs might result in
a highly spinning black hole from which the energy to launch two relativistic
radio jets can be extracted \citep{blandford77,chiaberge11}. At high redshift (z$>$1)
almost all radio galaxies are associated with recent or ongoing merger events
\citep{chiaberge15}. The properties of one such object, 3C~186, are
consistent with those expected if this is associated with a recoiling BH,
resulting from the coalescence of black holes in the late phases of a merger
\citep{chiaberge17}. Furthermore, there is some evidence  that mergers
also play a major role in radio loud objects at lower redshift
(e.g., \citealt{capetti06,baldi08,ramos11,ramos12}). Aside from the cases of 3C~75 and Abell~439, in which
two pairs of elliptical galaxies separated by $\sim$10-20 kpc are both
associated with an extended radio source \citep{owen85,cris85}, there have been
various claims of binary BHs in 3C sources. All of these assertions are based on the modeling of the parsec scale radio structure in these sources in which, however, only one black hole is active
\citep{depaolis04,lobanov05,roland13,romero00}. Two cases  deserve particular mention:
 0402+379,
the first binary black hole resolved as a visual binary system at 7.3 pc (\citealt{bansal17})
and NGC7675, which is to date the tightest AGN couple
ever imaged at a projected distance of 0.35 parsec (\citealt{kharb17}). Finally, no dual AGN are found
in the Chandra observations of the low $z$ 3C sample
\citep{massaro10,massaro12,massaro15}, but most of these observations were
taken with a short exposure time (8 ks).

We observed a sample of 20 radio galaxies with MUSE as part of the MURALES 
survey (a MUse RAdio Loud Emission line Snapshot survey). The sample
is formed by the 3C radio sources limited to $z<0.3$ and $\delta<20^\circ$,
visible during the April-September semester, i.e., R.A. $<$ 3$^{\rm h}$ and
R.A. $>$ 15$^{\rm h}$. The main aim of this survey is obtain deep line
emission images and explore the gas kinematics and its relationship with the
relativistic outflows. However, the sensitivity of MUSE combined with its the
high spatial and spectral resolution makes this instrument a powerful tool
to discover AGN pairs, as recently shown by \citet{husemann18}.

In this first paper we report the serendipitous discovery of a candidate dual
AGN in 3C~459, a radio galaxy at $z=0.220$ (where
$1\arcsec$ corresponds to $\sim$ 4.4 kpc). In radio, it is a luminous 
($L_{1400 \rm{MHz}}
= 2.1\times10^{25}$ W Hz$^{-1}$ sr$^{-1}$; \citealt{ulvestad85})  double-lobed source that is 37 kpc in size, is very
asymmetric with its eastern radio lobe brighter and closer to the nucleus,
and has an arm ratio of 5:1 \citep{thomasson03}. Both the lobes are edge-brightened.
The radio galaxy has also been detected in HI absorption with the Westerbork Radio Telescope (WSRT) \citep{morganti05}. 
In optical, there is some controversy about
the  classification of 3C~459. While the spectrum of
\citet{tadhunter02} does not show clear evidence for broad permitted lines, a
broad \ha\ component is instead detected by \citet{buttiglione09}. 
In X-ray, the hydrogen column density
 (N$_{\rm H} = 1.18 - 5.63 \times 10^{22}$ cm$^{-2}$;
\citealt{massaro12}) supports the identification of 3C 459 as a narrow-line galaxy.  On
the other hand, its radio core is relatively bright ($P_{\rm core}=1.6 \times
10^{33} \,\ergshz$ at 5 GHz), corresponding to a core dominance log$(P_{\rm
  core}/L_{178 \rm{MHz}})=-1.33$ typical of broad-lined radio galaxies
\citep{baldi13}.

The host of 3C~459 shows several signatures of a recent, gas-rich merger. A
large contribution of young stars emerges from the analysis of the optical
spectrum \citep{wills08,tadhunter11}, in line with the high star formation
rate ($\sim 100 M_\odot yr^{-1}$; \citealt{westhues16}) derived from modeling
its broadband spectral energy distribution. In deep, optical broadband images, the stellar component appears as highly
disturbed and have fans of diffuse emission that extend more than 30 kpc east and
south of the nucleus \citep{ramos11}.
In Sect. \ref{sample} we present the MUSE observations of 3C~459 and in
Sect. \ref{results} we present the main results, which are then discussed in
Sect. \ref{discussion}.

\begin{figure*}  
\centering{
\includegraphics[width=5.5cm]{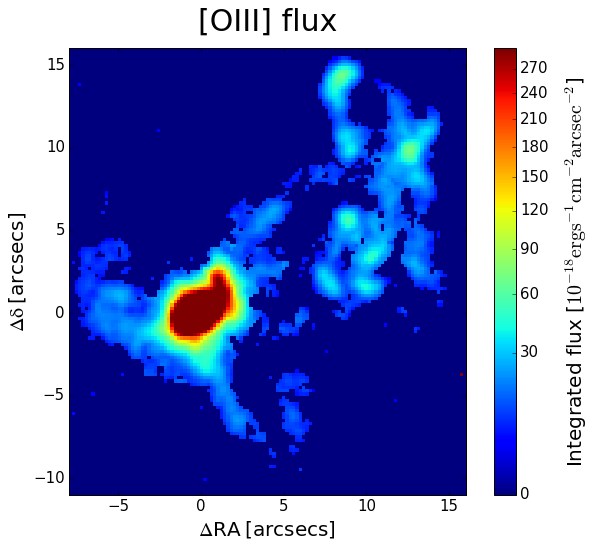}     
\includegraphics[width=5.55cm]{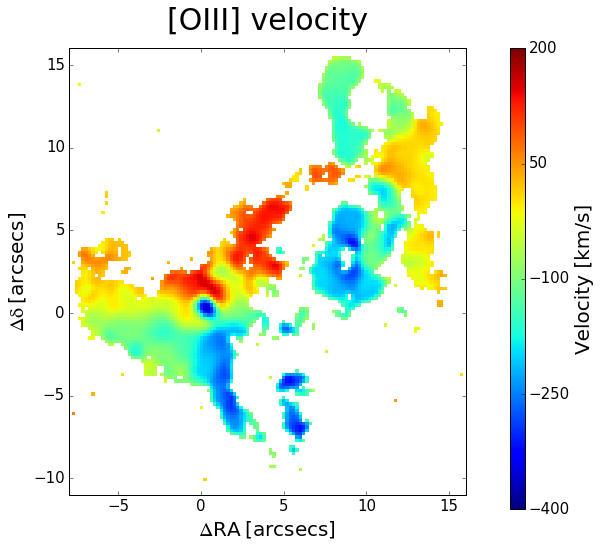}
\includegraphics[width=6.4cm]{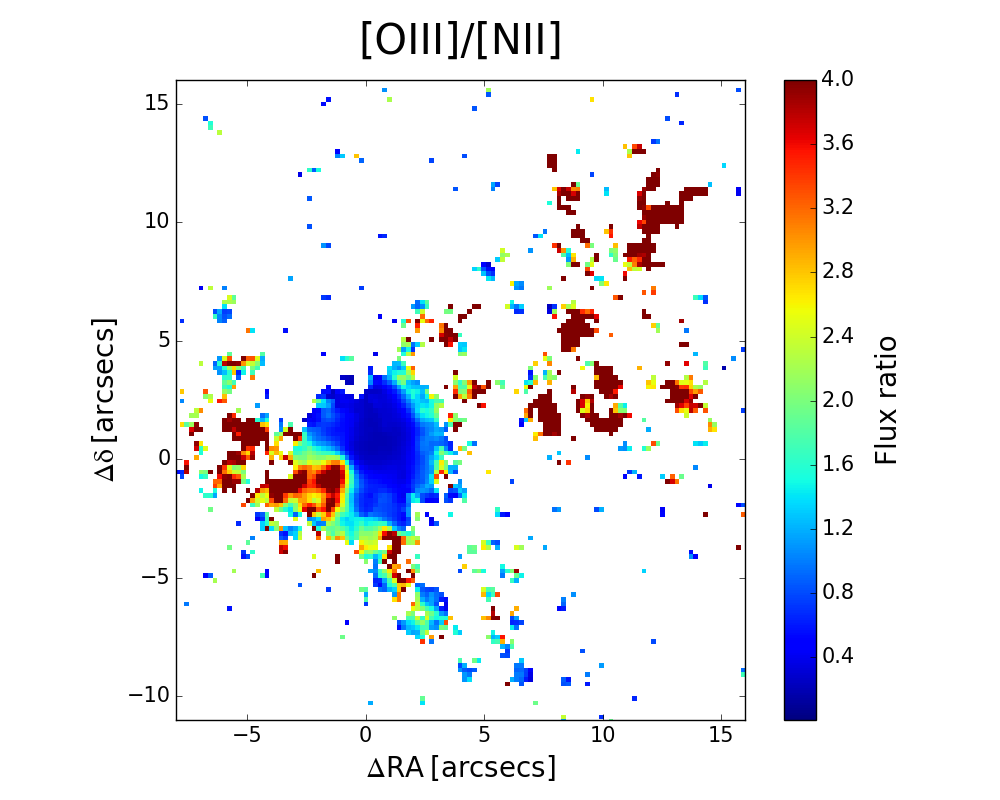}
\includegraphics[width=5.8cm]{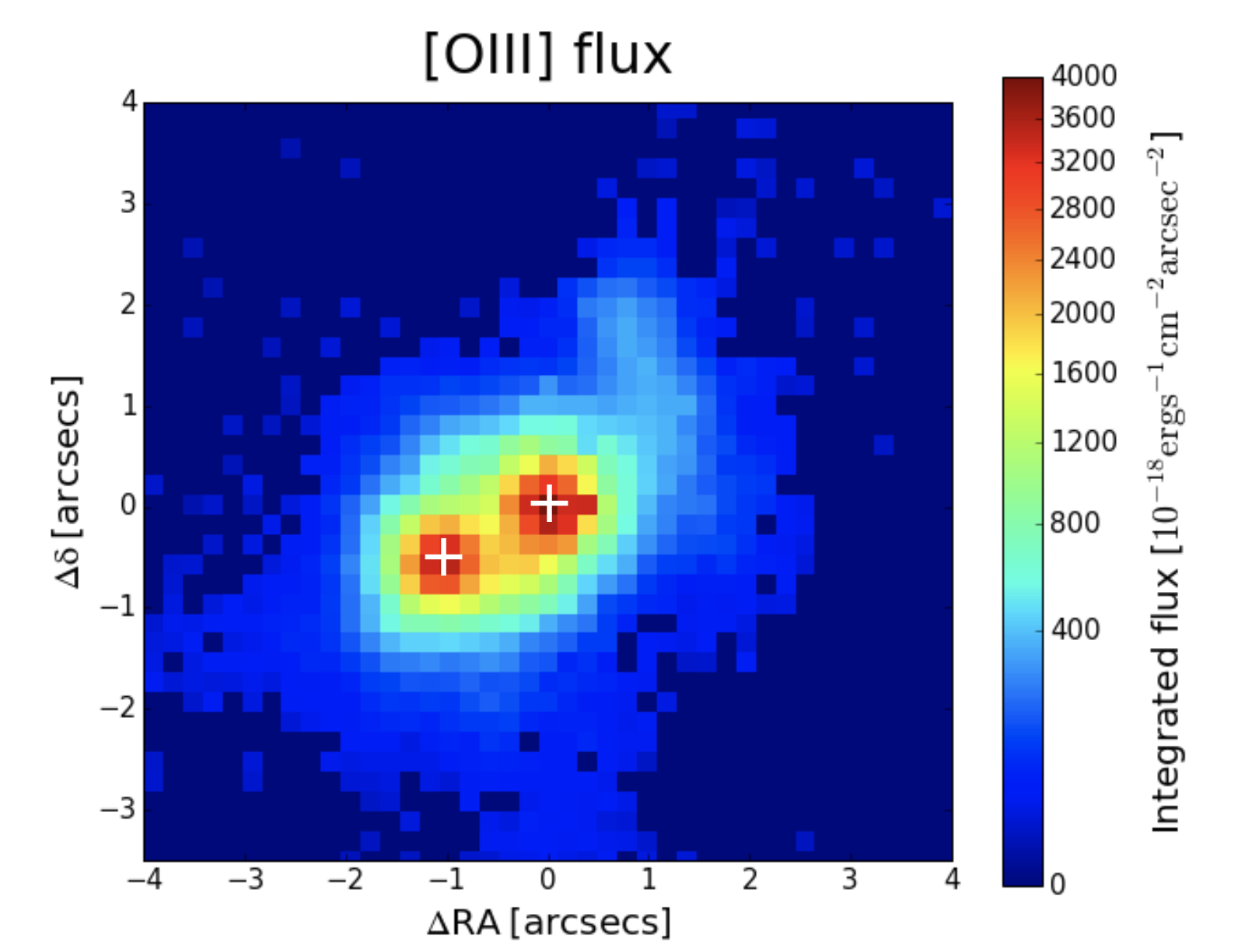} 
\includegraphics[width=5.8cm]{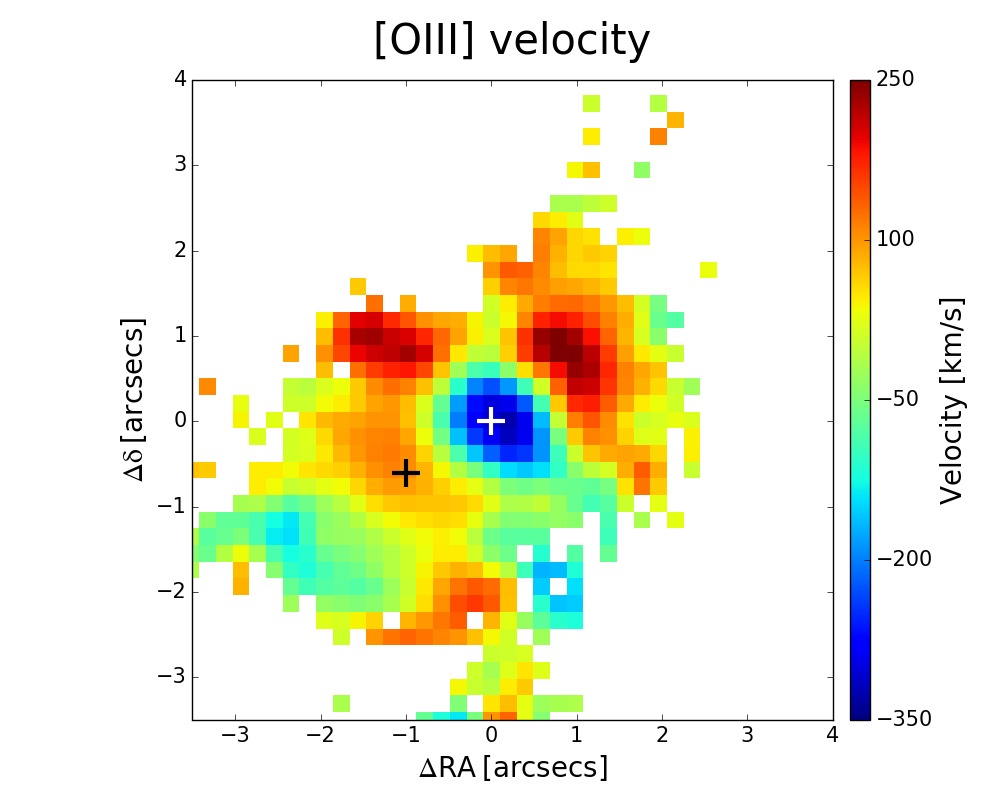}
\includegraphics[width=6.0cm]{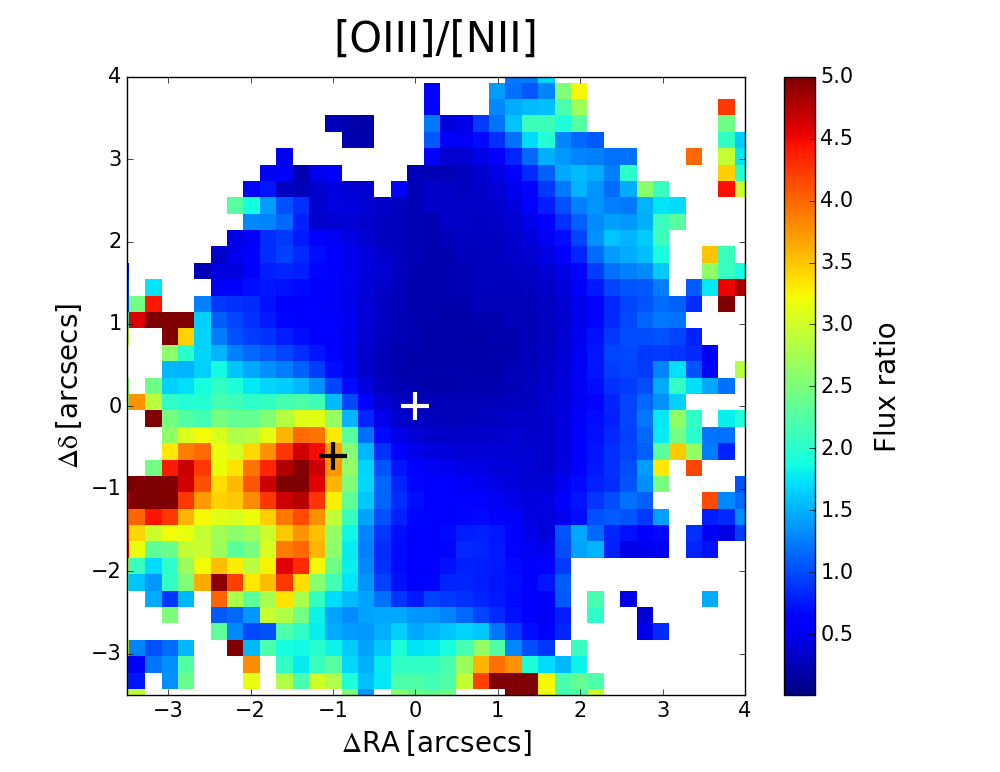}
\label{images}
\caption{Left: \oiii\ images smoothed with a Gaussian kernel of 0.5$\arcsec$
  covering the whole extent of the ionized gas nebula, 120 $\times$ 110 kpc,
  (top) and (bottom) zooming onto the central 33 $\times$ 33 kpc. Center:
  velocity field derived from the same line. Right: gas ionization map
  obtained from the ratio of the \oiii\ and \nii\ lines. The plus symbols indicate
  the location of $N1$ and $N2$.}}
\end{figure*}   

\begin{figure}  
\centering{
\includegraphics[width=9cm]{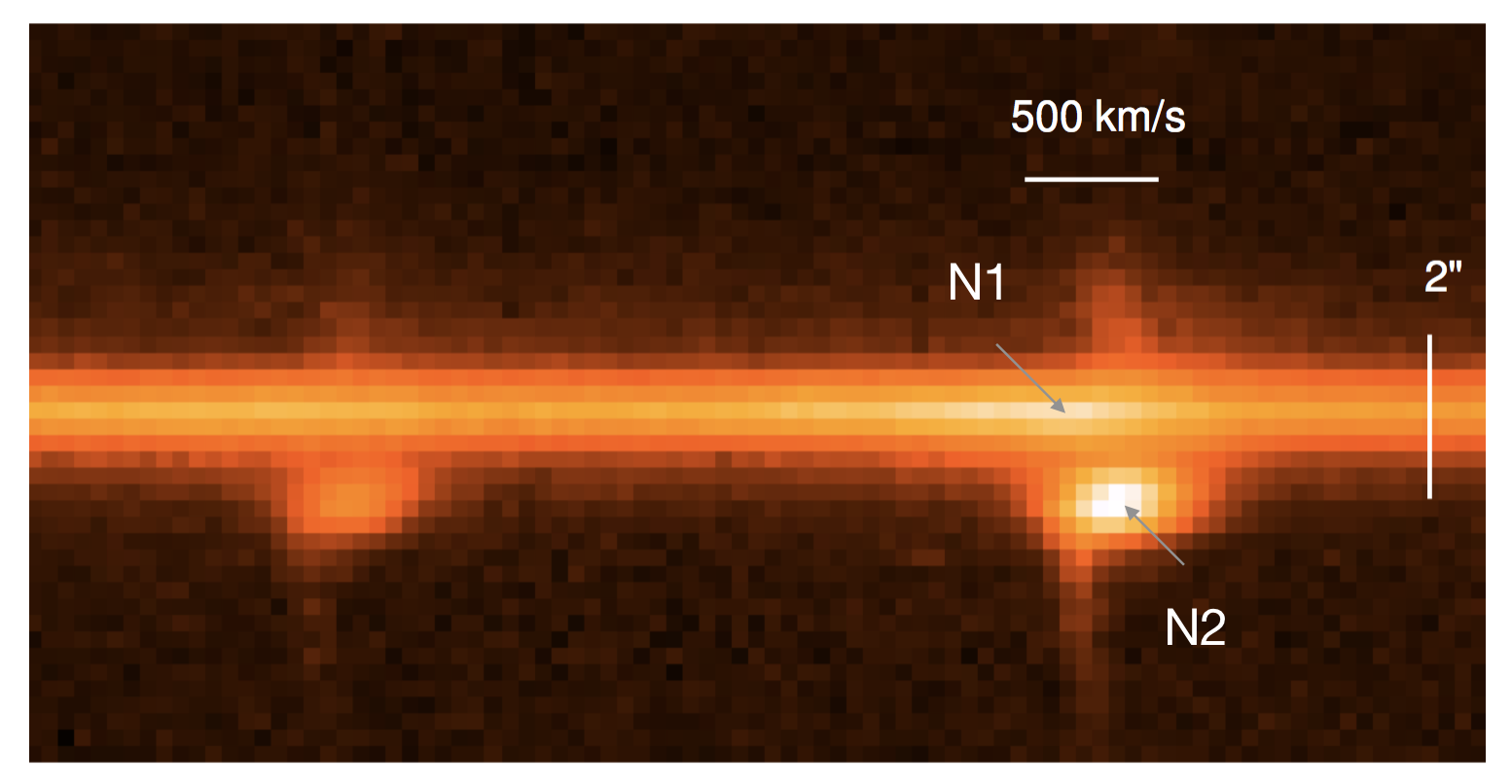}
\label{pv}
\caption{Position-velocity diagram for the [OIII] doublet extracted from a synthetic slit of 0.2$\arcsec$ along the direction joining the two nuclei.}}
\end{figure}

\begin{figure*}  
\centering{
\label{spectra}
\includegraphics[width=8cm]{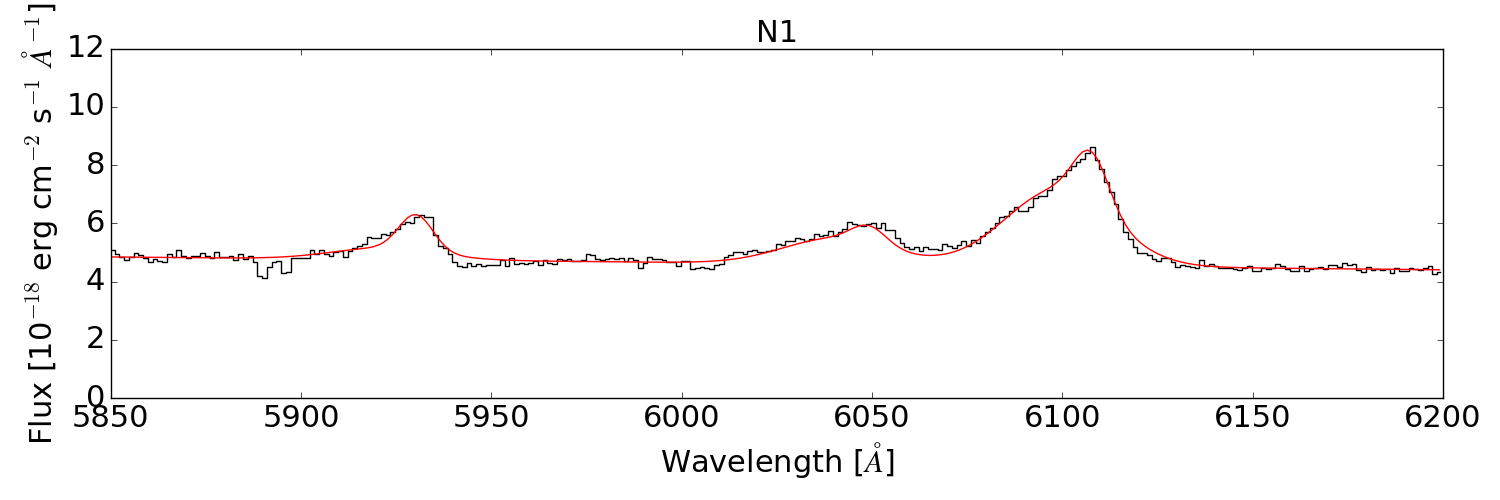}
\includegraphics[width=8cm]{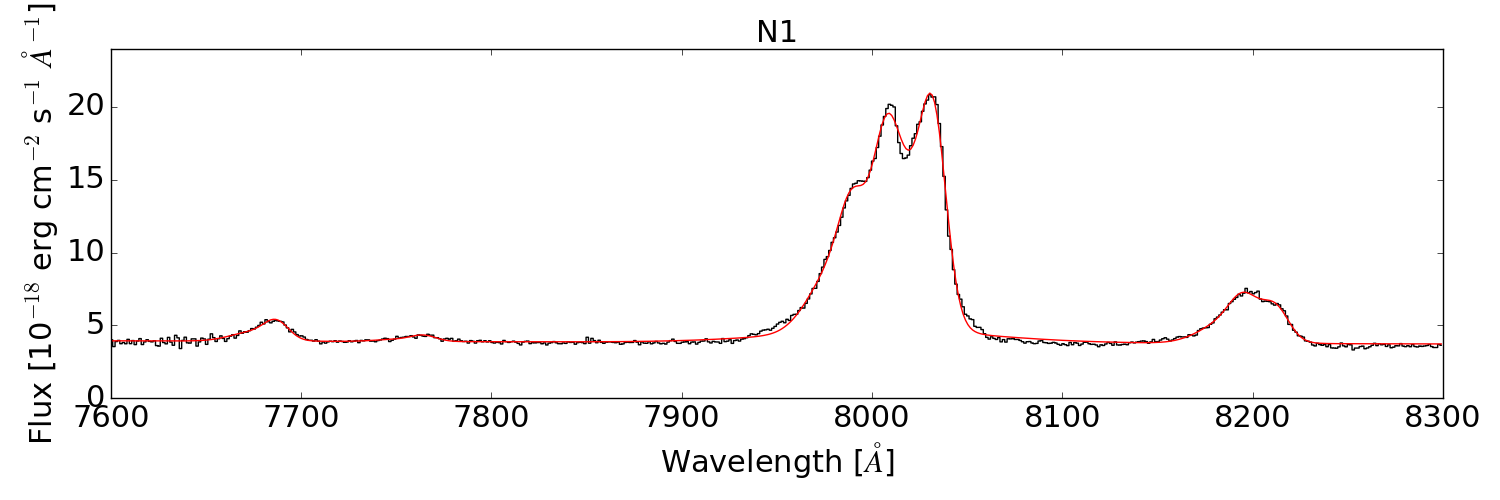}
\includegraphics[width=8cm]{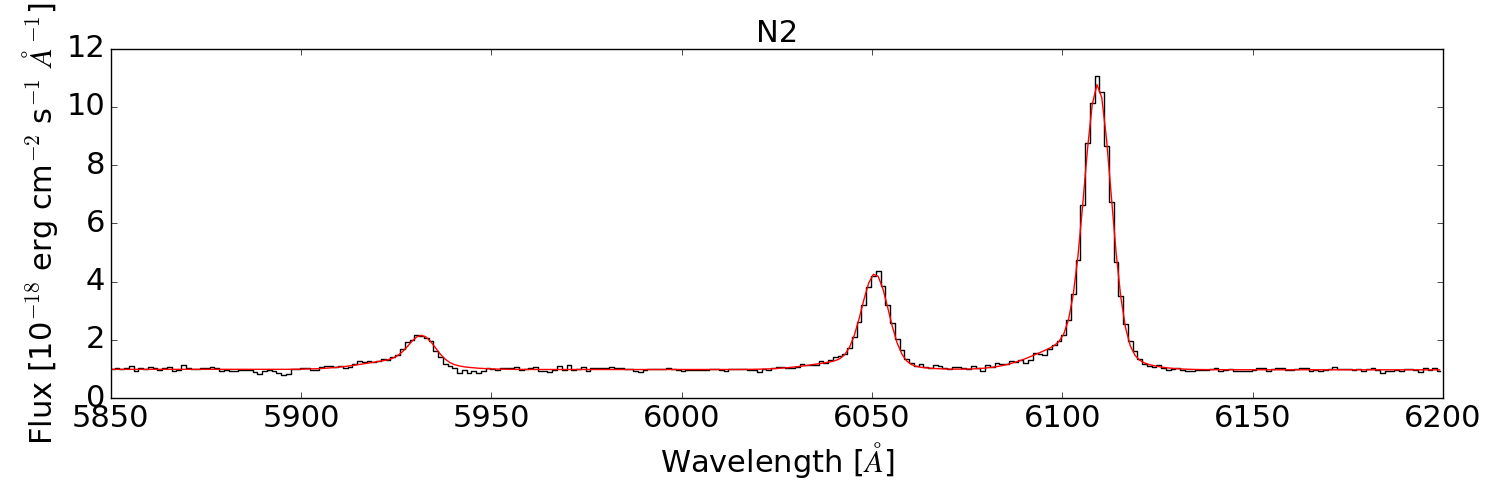}
\includegraphics[width=8cm]{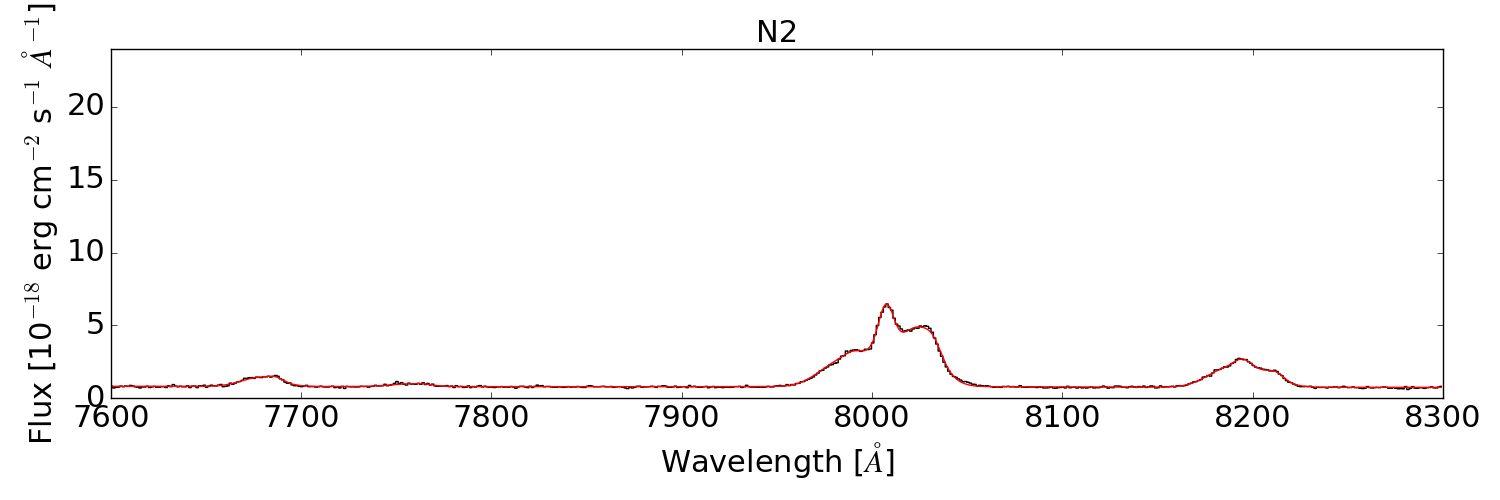} 
\caption{Comparison of the spectra of the two nuclear line knots (top for the
  nucleus of the radio galaxy, $N1$, bottom for the putative QSO2, $N2$) extracted
  from a synthetic aperture of diameter of 0$\farcs$6. Wavelengths are shown in
  observed frame.  In the spectra are visible the \hb\ and \oiii\ doublet
  on the left and \oi, the \nii+\ha\ complex, and the \sii\ doublet on the
  right.  }}
\end{figure*}

\section{Observation and data reduction}
\label{sample}

Two observations, with an exposure time of ten minutes each, were obtained with
the VLT/MUSE spectrograph on July 22, 2017 with a seeing of $\sim$
0\farcs5. We used the ESO MUSE pipeline (version 1.6.2) to obtain a fully
reduced and calibrated data cube.  

We followed the same strategy for the data analysis described in
\citet{balmaverde18}. Summarizing, we subtracted the stellar continuum after
resampling the data cube with the Voronoi adaptive spatial binning, requiring
an average signal-to-noise ratio per wavelength channel of at least 50. We
then used the penalized pixel-fitting code (pPXF; \citealt{cappellari03}) to
fit the absorption stellar features.

We  simultaneously fit all emission lines in the continuum subtracted spectra.
We assumed that all lines in the blue and red portion of the spectra have the
same profile, except for the \ha\ in the nuclear regions for which we allowed
for the presence of a broad component not required to fit the
\hb\ line. In the nuclear regions we included multiple Gaussian components,
while one component accurately reproduces the lines at larger radii.

\begin{table}
\label{tab0} 
\caption{Results of the spectral fit to the two nuclear sources, $N1$ and
  $N2$, extracted from a squared synthetic aperture of side 0\farcs6. Fluxes
  are in $10^{-16}$ erg cm$^{-2}$ s$^{-1}$ units. The relative flux errors are
  always smaller than 3\%.}
\centering                                    
\begin{tabular}{l r r}                     
\hline          
      
Line & Flux N1 & Flux N2 \\    
\hline
{[O~III]}$\lambda$5007  & 28.1  & 25.6   \\
H$\beta$                &  6.8  &  3.9   \\
{[OI]}$\lambda$6300     &  2.6  &  1.3   \\
H$\alpha$               & 60.1  & 20.2  \\
{[NII]}$\lambda$6584   & 113.7  & 26.9 \\
{[SII]}$\lambda$6731    & 20.3  & 7.0   \\
{[SII]}$\lambda$6717    & 12.9  & 8.7   \\
\hline          
\end{tabular}
\end{table}

\section{Results}
\label{results}

Line emission in 3C~459 is detected out to a distance of $\sim$18\arcsec\ (80
kpc) from the nucleus (see Fig. \ref{images}). On the NW side it is dominated
by several emitting filaments, confined within a triangular region with its
apex toward the center of the galaxy.

The central regions are dominated by two compact line emission peaks of
similar brightness: one (hereafter, we refer to it as $N1$) is located at the
position of the continuum peak and coincident also with the radio core and the
other ($N2$) is offset by $\sim1\farcs2$ (5.3 kpc) to the SE. The gas velocity
field (Fig. \ref{images}, central panels) is very complex; relative
velocities range from -350 to 250 \kms. Nonetheless, following in turn each
of the various NW filaments, we find rather smooth velocity gradients,
indicative of ordered motions. At the nucleus we also find a complex velocity
structure. In particular, there is a difference of $\sim$400 \kms\ between
the baricenter of the \oiii\ line between $N1$ and $N2$. The offset measured
at the lines peak is instead $\sim200$ km/s.

 We extracted a position-velocity
diagram for the \oiii\ line doublet  along the direction 
of the two nuclei (see Fig. 2) with a synthetic aperture of 0.2$\arcsec$. At large radii, we see 
gas extending over $\sim$6$\arcsec$ ($\sim$25 kpc) following a
rotational pattern. The two nuclei are both offset with respect to the center of rotation.

\begin{figure}  
\label{niiha}
\centering{
\includegraphics[width=9cm]{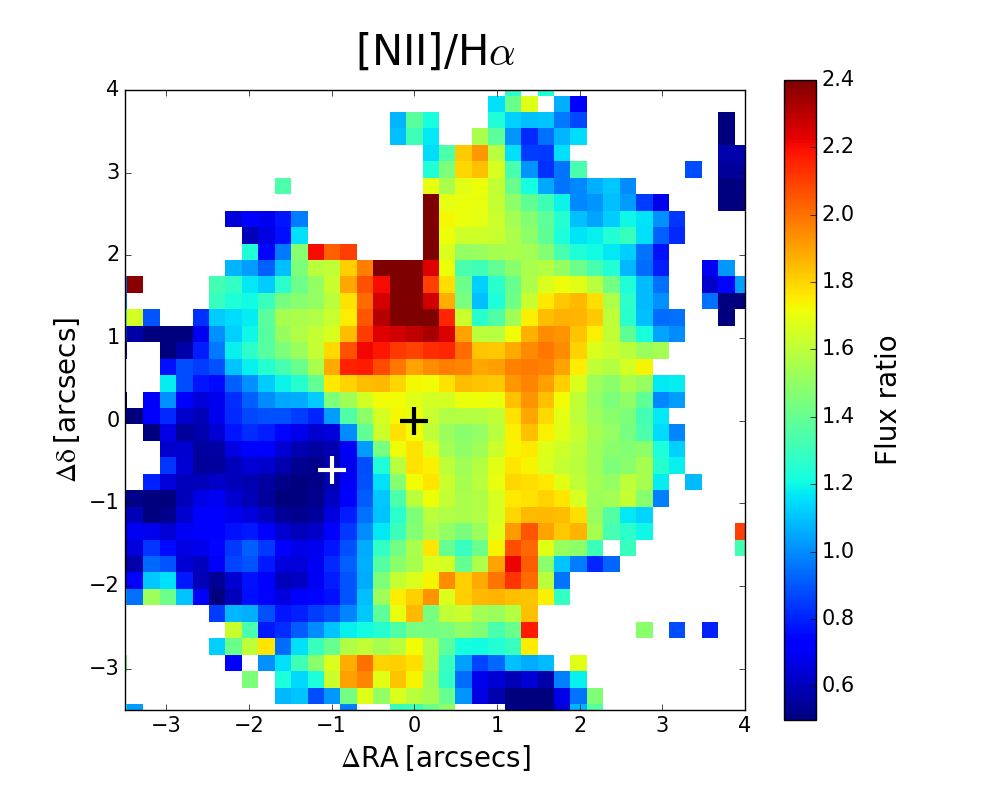}
\caption{\nii/\ha\ emission line ratio in the central 35
  $\times$ 35 kpc}}
\end{figure}

The $N1$ and $N2$ regions also show dramatic differences in their spectra. In Fig. \ref{niiha}  and
Table \ref{tab0},  we provide fluxes of the brightest emission lines
extracted from a squared synthetic aperture of side 0\farcs6. The lines are
broader on $N1$, which has line widths of $\sim$500 \kms, compared to $\sim$300
\kms\ measured on $N2$. In the spectroscopic diagnostic diagrams both regions
fall into the area populated by high excitation objects such as Seyferts, QSOs, and
high excitation radio galaxies. (\citealt{kewley06,buttiglione10}), but the
ratios of these objects are significantly different. In particular, the \oiii\ is relatively
stronger on $N2$ than on $N1$: the \oiii/\nii\ ratio is 0.25 on $N1$ and
0.58 on $N2$, indicating a higher gas ionization state.

These results suggest that $N2$ is associated with a second active
nucleus. The ratio \ha/\hb=5.1$\pm$0.1 (narrow components only) is indicative
of substantial internal reddening, amounting to E(B-V)=0.47 having adopted the
\citet{cardelli89} extinction law. The absorption corrected \oiii\ luminosity
of $N2$ is $9.2 \pm 0.6 \times 10^{41} \ergs$ because the error is dominated by
the uncertainty on the Balmer lines ratio. This luminosity value locates $N2$ well
within the range of luminous QSOs \citep{zakamska03,reyes08}.

Since no compact source is detected in the radio maps at the location of $N2$
(a rough upper limit of 2 mJy can be derived from the inspection of Fig. 3 of
\citealt{thomasson03}), we can identify this emitting region as a
radio quiet type 2 QSO based on the emission line ratios, the lack of
broad emission lines, and the \oiii\ luminosity.

In Fig. \ref{images}, right panels, we also present an image obtained by
dividing the \oiii\ and \nii\ images, a ratio image
sensitive to the ionization state of the emitting gas. In the central few
arcseconds from the radio nucleus the ratio is $\sim0.3-0.8$, represented by the blue patch
in this figure. On the SE side, however, this ratio map reveals the presence
of a triangular region of highly ionized gas with the apex located at $N2$. By
extending the boundaries of this region to the opposite side of the nucleus,
they include the large scale system of filaments seen to the NW. These
filaments are also characterized by a high \oiii/\nii\ ratio, similar to what
is seen on $N2$. The large spectral separation between the \oiii\ and
\nii\ lines opens the possibility that their ratio is due to changes in
reddening across the source. The \hb\ image is not of sufficient quality to
test directly this issue, but the \nii/\ha\ ratio map (not sensitive to
absorption) shows the same triangular structure seen in the \oiii/\nii\ map
(see Fig. 4).

This result strengthens the interpretation that $N2$ is indeed a type 2
nucleus. In these sources, the AGN radiation field is highly asymmetric due to
nuclear obscuration and, as a result, their narrow line region (NLR) often
show a biconical morphology (see, e.g., \citealt{tadhunter89,wilson93,williams17}). In
the case of 3C~459 the bicone partly overlaps with the NLR of the radio loud
AGN. The cartoon of Fig. 5 provides a schematic view of the
geometry of the system.

We looked for a signature of the presence of this secondary nucleus in other
bands. In Fig. 6 we collected the images from Chandra in X-rays, MERLIN in radio waves, and HST in both the optical and near-IR bands. No
emission is associated with $N2$ in any of these images. The only exception is the
F606W HST image, a broadband filter that includes the \oiii\ line, where at
the location of $N2$ there is an elongated feature, possibly due to line
emission. 

Based on the available information, we cannot rule out the possibility that
N2 is just a bright emission line knot resulting from, e.g., a jet-cloud
interaction, a process often observed in both Seyfert and radio galaxies
(e.g., \citealt{winge97,villarmartin99}). We do not favor this scenario based on a
list of reasons: 1) A jet-cloud interaction usually produces very broad line
profiles (e.g., \citealt{gelderman94,capetti99}); and conversely, even though the
separation of $\sim$400 \kms\ between $N1$ and $N2$  would require the
acceleration of the ionized to high velocities, the lines on $N2$ are narrower
than on $N1$ and symmetric; 2) Fig. 6 shows that $N2$ is not located along the
radio axis, but at the edges of the eastern radio lobe, $\sim30^\circ$ away
from the jet path;  3) the ionization structure centered on $N2$ would be a
fortuitous spatial coincidence; and 4) N2 has a very high line luminosity of
$\sim10^{42}$erg s$^{-1}$ that is typical of luminous QSOs.


\begin{figure}  
\centering{
\includegraphics[width=9cm]{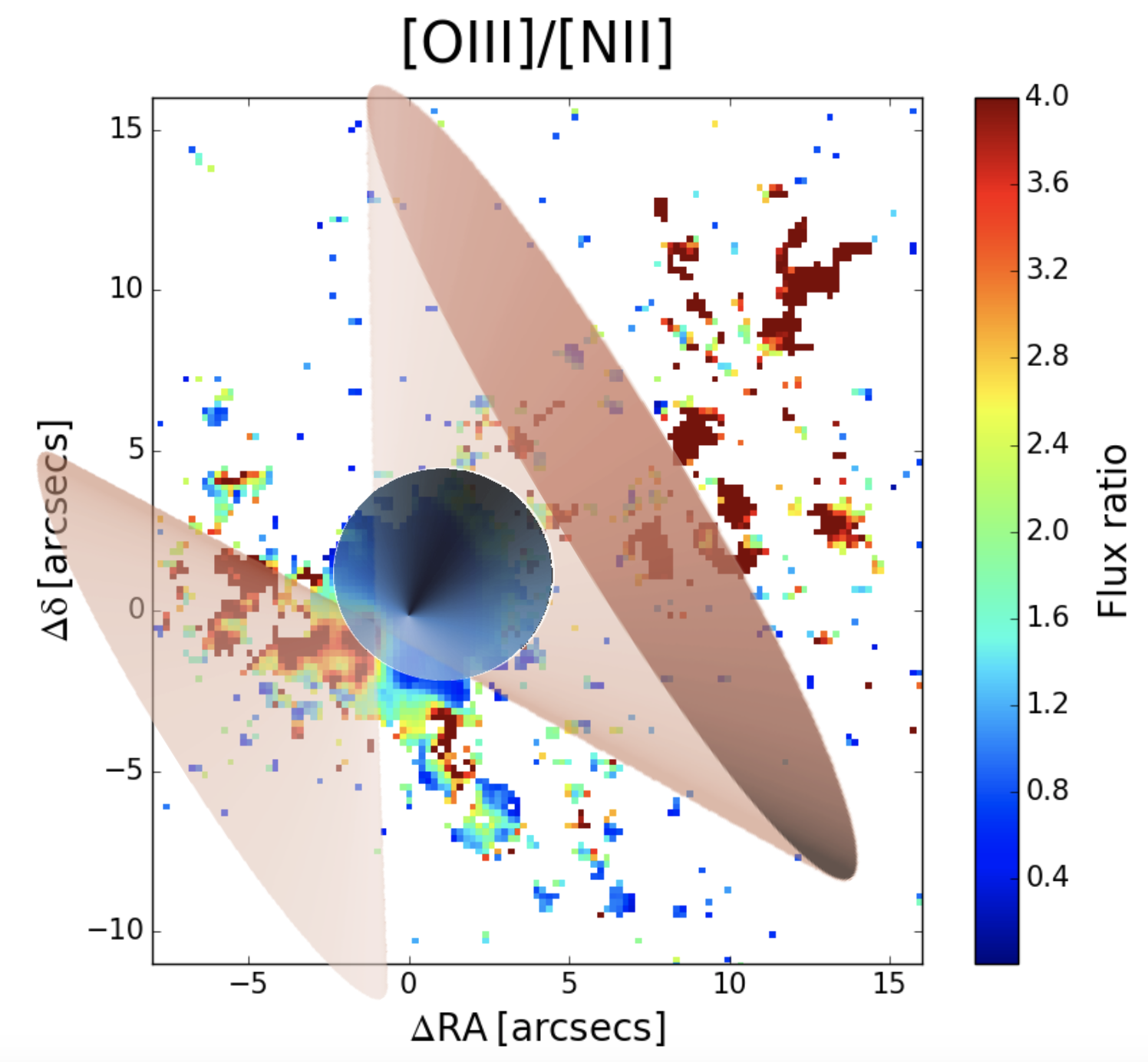}}
\label{cartoon}
\caption{Schematic representation of the geometry of the dual AGN in
  3C~459. The radio galaxy produces a low ionization halo, possibly a face on
  ionization cone, around the radio nucleus, while the hidden QSO nucleus
  generates a high ionization bicone extended for $\sim$88 kpc.}
\end{figure}   
  
\begin{figure}  
\centering{ \includegraphics[width=9cm]{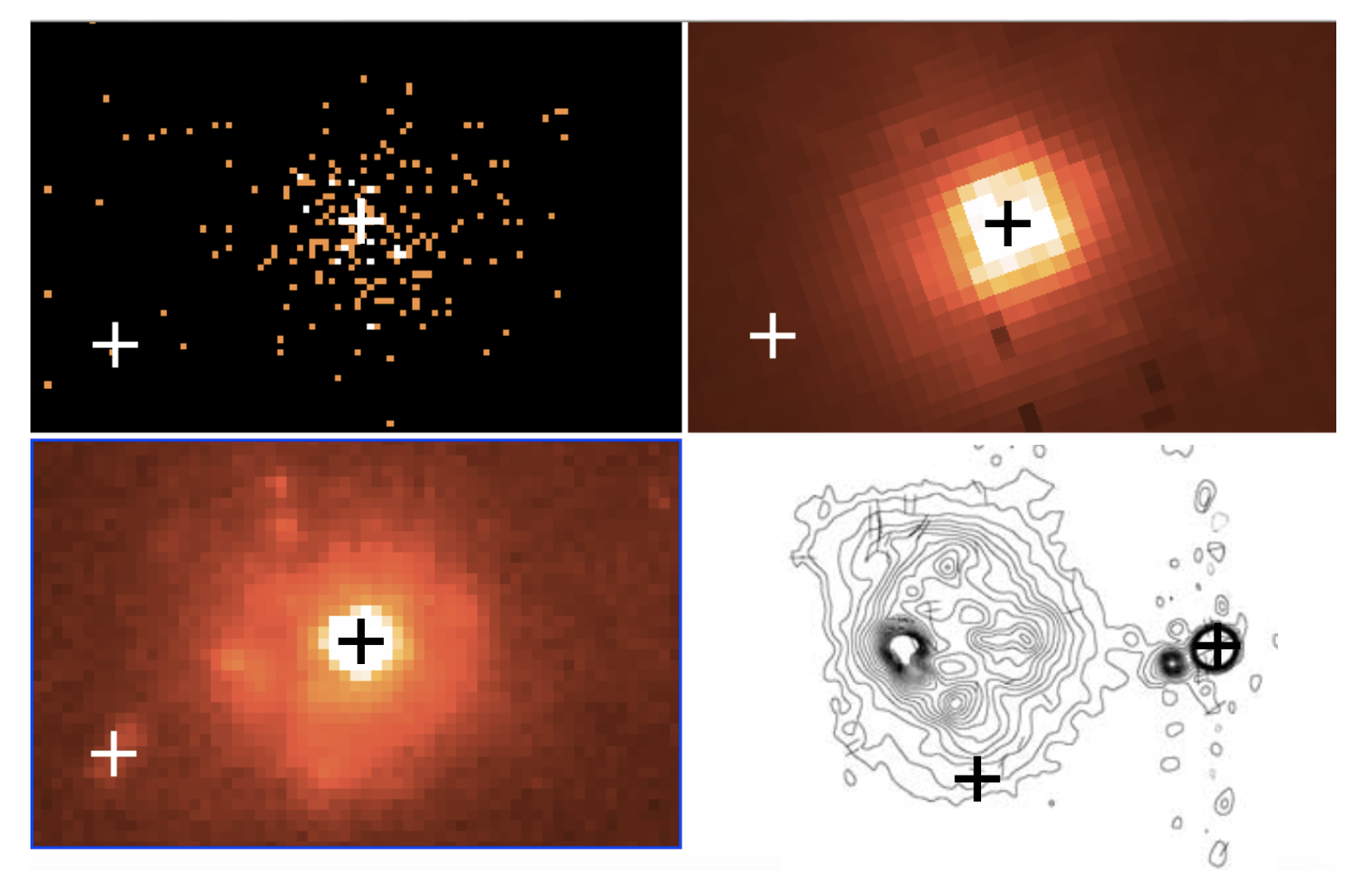}}
\label{nucleus}
\caption{Multiband images of the nuclear regions of 3C~459, Chandra (top left), infrared HST
  (top right), optical HST (bottom
  left),  and radio from MERLIN 1658 MHz (bottom right). The two crosses indicate the location of the emission line
  peaks. In the Chandra image the pixel 
size is 0.06$\arcsec$ (below the Chandra native resolution of 0.459$\arcsec$). 
The Merlin image has the same field of view, but it is shifted by 1$\arcsec$ to the west to show the radio structure best.
}
\end{figure}

\section{Discussion and conclusions}
\label{discussion}
The $N2$ source is a very bright object and has a \oiii\ luminosity of $\sim10^{42}$erg s$^{-1}$. The expected X-ray luminosity, adopting a standard $L_{\rm 2-10
  keV}/L_{\rm [O~III]}$ ratio for Seyfert I galaxies and QSOs
\citep{panessa06}, is $\sim7\times 10^{43} \ergs$. The detection of a single
high energy photon (above 5 keV) in the 60 ks Chandra image (see
Fig. 6) requires $N2$ to be a heavily absorbed, Compton thick
source. The expected radio emission is instead $\sim1$ mJy
\citep{ho01,panessa07}. This low flux level makes a radio detection very
challenging, also considering that it would be embedded in the western lobe of
3C~459. A similar flux density is predicted in the mid-infrared window
\citep{horst08,gonzalez13}. This is well below the sensitivity of current
instruments (e.g., \citealt{asmus14}), but easily reached with the James Webb
Space Telescope. An alternative approach to explore the nature of $N2$
further is to employ ALMA\ for a high resolution search for the nuclear molecular gas expected to be
associated with the secondary AGN \citep{villar13}.

3C~459 is the only dual AGN candidate emerging from a preliminary analysis of
the MURALES data obtained so far, that includes 20 radio galaxies, 15 of which
are FR~II. However, it might not be an isolated case. \citet{tremblay09}
presented the results of a project of emission line imaging of the 3C sample
with HST, also limited to $z<0.3$: data are available for 19 objects, only two
in common with MURALES. Two galaxies, namely 3C~136.1 and 3C~196.1, show a
\oiii\ morphology similar to 3C~459, which have two compact regions of line emission
separated by $\sim0\farcs5$, corresponding to $\sim$0.5 and 1.5 kpc,
respectively \citep{tremblay09}. Although all
these sources require further analysis, the fraction of dual AGN in the
3C sample could be significant, on the order of $\sim$10\%. Improved
insight into this phenomenon, and  its connection with the dynamical stage of
the mergers, would be of great importance to explore the role of mergers in
the triggering and in evolution of powerful radio sources. Further MUSE
and HST observations are needed to set these results on stronger statistical
grounds.

\begin{acknowledgements}
Based on observations made with ESO Telescopes at the La Silla Paranal
Observatory under program ID 097.B-0766(A). This research has made use of
data obtained from the Chandra Data Archive. The National Radio Astronomy
Observatory is a facility of the National Science Foundation operated under
cooperative agreement by Associated Universities, Inc. B.B. acknowledge financial contribution from the
agreement ASI-INAF I/037/12/0. Some of the data presented are based on observations made with the NASA/ESA Hubble Space Telescope, obtained from the data archive at the Space Telescope Science Institute. STScI is operated by the Association of Universities for Research in Astronomy, Inc. under NASA contract NAS 5-26555. We thank the referee for her/his comments.
\end{acknowledgements}


\end{document}